\documentclass[aps,pra,twocolumn,amsmath,amssymb, superscriptaddress,tightenlines]{revtex4}
\usepackage{graphicx}
\usepackage{epsfig}
\usepackage{dcolumn}
\usepackage{bm}
\usepackage{amssymb}
\usepackage{bm}
\usepackage{amsmath, bm}
\usepackage{upgreek}
\usepackage[colorlinks,citecolor=blue,linkcolor=blue,hyperindex]{hyperref}

\begin{document}

\title{Comment on ``Interpretation of thermal conductance of the $\nu=5/2$ edge''}
\author{D. E. Feldman}
\affiliation{Department of Physics, Brown University, Providence, Rhode Island 02912, USA}
\date{\today}



\begin{abstract}
We address the interpretation proposed in the paper [Simon, Phys. Rev. B {\bf 97}, 121406(R) (2018)] of the thermal conductance data from [Banerjee {\it et al.}, Nature {\bf 559}, 205 (2018)]. We show that the interpretation is inconsistent with experimental data and the sample structure. In particular, the paper misses the momentum mismatch between contra-propagating modes. Contrary to the claim of the paper, low energy tunneling involves a large momentum change. We consider only the ``small Majorana velocity'' mechanism [Simon, Phys. Rev. B {\bf 97}, 121406(R) (2018)]. Other mechanisms, interpretations of the experiment, and their difficulties are beyond the scope of this Comment.
\end{abstract}

\maketitle

For a long time, there has been tension between numerical and experimental findings about the quantum Hall effect at $\nu=5/2$ in GaAs. A seminal paper \cite{x1-Morf} by Morf established the non-Abelian Pfaffian state \cite{x2-MR} as a viable possibility. The numerical results \cite{x1-Morf} were later reinterpreted as supporting also the anti-Pfaffian topological order \cite{x3-LRH,x4-LRNF}.
At the same time, numerical results for the energy gap \cite{x5-Morf,x6-PRX} have remained many times higher \cite{x5-Morf} than even the highest experimentally measured gap \cite{x7-Csathy}. Tunneling experiments were interpreted \cite{x8-XiLin-2012,x9-Baer} as supporting the Abelian 331 \cite{x10-Halperin} and 113 \cite{x11-YF-2014} states in an apparent conflict with numerics. More recently, it was proposed \cite{x12-ZF} that the existing experimental data can be explained by the 
non-Abelian PH-Pfaffian order \cite{x4-LRNF,x18-Son,x18a-Ashwin,x18b-Bonderson}. The PH-Pfaffian hypothesis implies two predictions for the thermal conductance. First, the heat conductance of a large sample must be quantized at $KT=2.5\kappa_0 T$, where $\kappa_0 T=\pi^2k_B^2 T/3h$ is one thermal conductance quantum \cite{x12-ZF,x18-Son}. Second, a peculiar edge structure of the PH-Pfaffian liquid implies unusually rapid growth of $K$ above 
the universal quantized value at sufficiently low temperatures \cite{x19-Nature-2018} (Methods section). Both predictions are consistent with the results of a recent experiment \cite{x19-Nature-2018}.
 
Yet, the interpretation of the thermal conductance data is tricky. Indeed, assuming that the central Ohmic reservoir \cite{x19-Nature-2018} is in thermal equilibrium, the observed thermal conductance should be understood as an upper bound on the universal theoretical heat conductance along the edges of a large sample. One reason is the coexistence of edge and bulk heat transport. In particular, the measured heat conductance includes a phonon contribution. That contribution can be subtracted with an ingenious trick \cite{x21-Science-2013}. Still, even the edge heat conductance can exceed the universal quantized value. Indeed, the thermal conductance of a long edge equals the difference of the heat conductances of the upstream and downstream edge modes \cite{x22-KF-eq}. In a very short sample, the contributions of the upstream and downstream modes add up. The observed heat conductance can farther increase due to edge reconstruction, which creates additional up- and down-stream modes in a short sample \cite{x23-CW}. Naturally, all intermediate values between the sum and difference of the upstream and downstream heat conductances might also be observed, depending on the sample details. 

Fortunately, the same experimental techniques as at $\nu=5/2$ can be used to measure the thermal conductance at better understood filling factors. It turns out that the measured heat conductance at the filling factors $1/3,~4/7,~3/5,~2/3,~7/3$, and $8/3$ is consistent with the theoretical predictions for a long sample \cite{x19-Nature-2018,x24-Nature-2017}. The greatest difference of the theoretical and experimental thermal conductances was observed at $\nu=2/3$, but even there the observed $KT=0.25\kappa_0T~-~0.33\kappa_0 T$ is much closer to the universal value of $0$ than to the sum of the upstream and downstream heat conductances $2\kappa_0T$. Note also that the equilibration length is similar at $\nu=2/3$ and other filling factors \cite{x24-Nature-2017}. A greater deviation from the theoretical value of the heat conductance is due to an unusually slow dependence \cite{x24-Nature-2017} of the thermal conductance  on the sample length at $\nu=2/3$. Based on the results at multiple filling factors, it is thus natural to conclude that the measured $KT\approx 2.5\kappa_0T$ is close to the universal quantized heat conductance of the $5/2$ liquid, in agreement with the PH-Pfaffian hypothesis.

An interesting recent paper \cite{x25-Simon} challenges this conclusion. It suggests that due to a slow velocity of the Majorana mode and the smoothness of the random potential in the sample, an upstream Majorana mode fails to equilibrate with the rest of the edge modes. Then the observed $K$ might be compatible with the anti-Pfaffian state.

The goal of this comment is to analyze the assumptions behind the physical picture \cite{x25-Simon}. We find that they are incompatible with the existing experimental data and the sample structure. In particular, Ref. \onlinecite{x25-Simon} misses the momentum mismatch between contra-propagating modes. Contrary to the claim of the paper, low energy tunneling involves a large momentum change. Our only focus is the ``small Majorana velocity" mechanism \cite{x25-Simon}. A detailed discussion of other mechanisms and interpretations is beyond the scope of the Comment. 

The edge-equilibration picture assumed in Ref. \onlinecite{x25-Simon} differs from the classical Kane-Fisher-Polchiski picture \cite{x26-KFP} of the equilibration in the disorder-dominated regime, as extended to $\nu=5/2$ in Refs. \onlinecite{x3-LRH} and \onlinecite{x4-LRNF}. We will start with a brief review of the classical picture. This will help us translate the qualitative language of Ref. \onlinecite{x25-Simon} into equations. Such translation is necessary to understand when the picture \cite{x25-Simon} is applicable. We will discover that it does not apply to the sample from the experiment \cite{x19-Nature-2018}. 

The anti-Pfaffian edge includes two downstream integer channels, an additional downstream charged Bose-mode $\phi_1$ with the conductance $e^2/h$, an upstream Bose-mode $\phi_2$ with the conductance $e^2/2h$, and an upstream neutral Majorana $\psi$. There is Coulomb interaction between charged modes and electron tunneling between various channels. We summarize this picture with the following action, which omits integer modes for brevity:

\begin{eqnarray}
\label{1}
L=\frac{1}{4\pi}\int dx dt [\partial_x\phi_1(\partial_t\phi_1-v_1\partial_x\phi_1)-2\partial_x\phi_2(\partial_t\phi_2+v_2\partial_x\phi_2) \nonumber\\
-2v_{12}\partial_x\phi_1\partial_x\phi_2]
+\int dx dt [i\psi(\partial_t\psi+u\partial_x\psi)]+L_{\rm tun},
\end{eqnarray}
where $v_1$, $v_2$, and $u$ are the mode velocities, $v_{12}$ is the Coulomb interaction, and $L_{\rm tun}$ describes tunneling with a random complex amplitude $W(x)$:

\begin{equation}
\label{2}
L_{\rm tun}=\int dx dt [W(x)\psi\exp(2i\phi_2+i\phi_1)+{\rm H.c.}].
\end{equation}
The tunneling amplitude $W(x)$ is assumed to have only short-range correlations. The tunneling term is responsible for charge equilibration and the observed quantized electrical conductance. Tunneling is relevant in the renormalization group sense in a strongly interacting system. It is also likely strong since the unscreened random potential is expected to exceed the energy gap \cite{x27-MAH}. Hence, the problem is strongly coupled in the language of the modes $\phi_{1,2}$ and $\psi$. The weak coupling description is possible on large scales in the language of a single downstream charged boson and three emergent upstream Majorana modes that propagate with the same speed \cite{x3-LRH,x4-LRNF}. Random tunneling $L_{\rm tun}$ disappears in that language and weak residual random inter-mode interaction is responsible for the energy equilibration between upstream and downstream modes. In the absence of the energy equilibration, the heat conductance is the sum of $3$ quanta from the Bose modes and $1.5$ quanta from Majoranas. This exceeds the observed thermal conductance of $2.5\kappa_0T$.

Ref. \onlinecite{x25-Simon} proposes to use the language of Eqs. (\ref{1},\ref{2}) at all length scales. This implies the assumption of weak $W(x)$. 
In this picture, the effect of $W(x)$ can be described in terms of energy and momentum conservation in scattering for elementary excitations of the Bose and Majorana modes [we ignore $v_{12}$ below; this does not significantly affect the argument]. Ref. \onlinecite{x25-Simon} assumes that the correlation length of $W(x)$ is large and hence the random potential supplies low momentum in scattering events. Finally, it is assumed that the Majorana mode is considerably slower than the Bose modes. Based on Ref. \onlinecite{x28-HRWY-2009}, $u$ is estimated as $u\sim 10^6~ 
{\rm cm/s}\sim v_{1,2}/6~-~v_{1,2}/8$. 

Let $\Delta k_{1,2}$ be the momentum changes of the modes $\phi_{1,2}$ in a scattering event. The momentum change of the Majorana mode is then $q-(\Delta k_1+\Delta k_2)$, where the momentum $q$ is supplied by $W(x)$. The energy conservation reads:

\begin{equation}
\label{3}
v_1\Delta k_1 - v_2\Delta k_2 - u(q-\Delta k_1 -\Delta k_2)=0.
\end{equation}
The equilibration process involves energy exchange between the Bose modes with $\hbar v_{1,2}\Delta k_{1,2}\sim k_{\rm B}T$. Ref. \onlinecite{x25-Simon} implies that $\hbar u(q-\Delta k_1-\Delta k_2)\ll k_{\rm B}T$ so that the Majorana mode does not equilibrate. Since $u\ll v_{1,2}$ this is equivalent to $\hbar uq\ll k_{\rm B} T$. Thus, the maximal momentum due to $W(x)$ must satisfy 

\begin{equation}
\label{4}
q_{\rm max}\ll k_{\rm B}T/\hbar u. 
\end{equation}

We are now ready to compare the above physical picture with what is known about the sample \cite{x19-Nature-2018}.

1) Let us substitute $T\sim 10~ {\rm mK}$ and $u\sim 10^6~{\rm cm/s}$ in Eq. (\ref{4}). We discover that $W(x)$ must be smooth on the scales $d\sim 1/q_{\rm max}$ on the order of microns.
At the same time, the distance between the 2D electron gas and the remote ionized dopants in the sample \cite{x19-Nature-2018} is 85 nm. That distance sets the maximal momentum supplied in scattering off an impurity. It is two orders of magnitude too high for the picture \cite{x25-Simon} to apply. Note also that a small number of impurities are inside the 2D gas. 
Ref. \cite{x25-Simon} correctly observes that disorder is often assumed to be a relatively long wavelength in high-mobility heterostructures. At the same time, this assumption is normally made about disorder far from the edges, and the disorder wavelength is assumed to be on the order of the setback distance to the remote ionized impurities 
\cite{x44-d2,xnew-review}.

2) Moreover, it would not save the mechanism \cite{x25-Simon}, if the random potential were smooth on the scale of microns \cite{footnote2} as the potential due to etched trenches might be. Indeed, if the disorder were so smooth, an edge of the length of microns would see a translationally invariant impurity potential. Yet, $W(x)$ would not be constant on such length scales. Indeed, besides the disorder effect, the amplitude $W(x)$ contains information about the momentum mismatch between different modes \cite{x3-LRH}. Eq. (\ref{2}) describes electron tunneling in the presence of a strong magnetic field. As is well known in the theory of momentum resolved tunneling in translationally invariant systems (see, e.g., Refs. \onlinecite{x29-Kun,x30-YF-2010}), $W(x)\sim\exp(i\Delta k x)$, where the momentum mismatch $\Delta k$ is set by microscopic length-scales of the problem. 
The anti-Pfaffian state can be understood as the Pfaffian state of holes inside an integer quantum Hall liquid. $\Delta k$ is proportional to the distance $s$ between the outer integer edge and the inner fractional edge.
We thus have to substitute $q=\Delta k+\delta k$ in Eq. (\ref{3}),
where $\delta k\ll\Delta k$ is a small contribution due to the spatial variation of the impurity potential. We then find that $\hbar uq\approx \hbar u\Delta k\gg k_{B} T$ and the system cannot equilibrate at all. In particular, the rapidly oscillating amplitude $W(x)$ disappears under the action of renormalization group at the thermal length scale $\sim \hbar u/k_{\rm B}T$. Hence, there should be no tunneling between the modes at low temperatures and $K=4.5\kappa_0$.

3) The inequality $\hbar u\Delta k\gg k_{B} T$ would not hold for a very low $u$ or $\Delta k$. Then the equilibration process would involve Majorana excitations with the momenta $q\sim\Delta k$. Hence, the mechanism \cite{x25-Simon} would require that

\begin{equation}
\label{5}
\Delta k\ll k_{\rm B}T/\hbar u.
\end{equation}
We expect \cite{zuelicke} that the momentum mismatch $\Delta k=s/l_m^2$ is not much lower than $1/l_m$, where $l_m\approx 10~{\rm nm}$ is the magnetic length. The velocity $u\sim 10^6~{\rm cm/s}$ is three orders of magnitude too high for Eq. (\ref{5}) to hold \cite{xnew-foot2}.

4) What if for some unlikely unknown reason $\Delta k\ll 1/l_m$? The mechanism \cite{x25-Simon} still does not work since then our starting point, Eq. (\ref{1}), is no longer valid. Indeed, besides random tunneling between the modes, the action also contains nonrandom tunneling \cite{x3-LRH}. At a large momentum mismatch, weak nonrandom tunneling can be neglected \cite{x3-LRH}, as seen, e.g., from the perturbative renormalization group analysis of rapidly oscillating contributions to the action. The assumption of small $\Delta k$ implies strong nonrandom tunneling since $\Delta k$ is proportional to the distance between the modes. Moreover, the nonrandom tunneling amplitude does not oscillate on the thermal length scale $\hbar u/k_B T$. Hence, nonrandom tunneling cannot be neglected. The analysis, based on Eq. (\ref{1}), does not apply \cite{xnew-foot3}.

5) At $\nu=5/2$, $K$ exhibits a dramatically stronger temperature dependence than at $\nu=2/3$. In contrast to the PH-Pfaffian hypothesis, the picture \cite{x25-Simon} does not explain the unusually rapid growth of $K$ at the lowest probed temperatures at $\nu=5/2$. 

The above points apply to the scenario \cite{x25-Simon} and do not mean that a different scenario of partial edge equilibration is impossible. A scenario, free from issues 1)-4), was proposed in Ref.  \onlinecite{x33-Harvard-talk}. The basic idea is in a sense opposite to Ref. \onlinecite{x25-Simon}: The classical picture of 
edge equilibration \cite{x3-LRH,x4-LRNF,x26-KFP} is used; it is observed that the charged mode of conductance $5e^2/2h$ in a system of integer and fractional channels may be much faster than the rest of the modes;  some additional assumptions are made.
The mechanism \cite{x33-Harvard-talk} faces its own challenges, but this comment is not an appropriate venue for their discussion or a discussion of any other mechanisms than \cite{x25-Simon} and their difficulties. We address the mechanism \cite{x33-Harvard-talk} elsewhere.

Here we limit our discussion of other mechanisms to a modified version of the mechanism \cite{x25-Simon}. The modification solves challenges 1)-4) but brings new issues. 
What if the Majorana mode is so slow that its thermal length (\ref{4}) remains much shorter than 10 nm even at $T\sim 10~{\rm mK}$? In other words, what if $u<10^3~{\rm cm/s}$?

6) The first new issue is the lack of evidence for such slow edge velocities. The only relevant numerical data \cite{x28-HRWY-2009} suggest $u\sim 10^6~{\rm cm/s}$. This is comparable with the existing experimental data \cite{x34-neutral-speed} for the neutral mode velocity at $\nu=2$. This is also consistent with the simplest theoretical estimate \cite{x11-YF-2014} of the edge mode velocity $\sim e^2/\epsilon h$, where $\epsilon$ is the dielectric constant. 

At the same time, many scenarios for the neutral mode velocity exist \cite{x11-YF-2014}, and it is essential to see what the existing data at $\nu=5/2$ imply for $u$. This leads us to

7) Current oscillations were reported \cite{x35-int1,x36-int2,x37-int3} in interferometers at $\nu=5/2$. According to Ref. \onlinecite{x38-int-theor}, the observed
 \cite{x36-int2,x37-int3} phase jumps by $\approx\pi$ reflect changes in the number of Majorana fermions in the interferometer. Such effect can only be observed if the thermal length is not much shorter than the micron-size interferometer \cite{x28-HRWY-2009}. This excludes $u<10^3~{\rm cm/s}$.

8) The assumption of $u<10^3~{\rm cm/s}$ implies that the energy scale $E_M$ of neutral-mode excitations is much lower than 10 mK even at the length scale of $l_m\approx 10~{\rm nm}$. 
The effective hydrodynamic model (\ref{1}) does not apply at the shorter scales $1/q>1/l_m$ and can only be used to describe low energy transport of the excitations with the energies $\epsilon(q)<E_M$. The energy flux, carried by such excitations, is not quantized at $\kappa_0 T^2/4$ in contradiction with a basic assumption of the picture \cite{x25-Simon}. The model (\ref{1}) cannot be used to find the contribution of higher-energy Fermi excitations. At a low $E_M$, it is plausible that the bulk gap for neutral excitations is not much greater than the experimentally relevant \cite{x19-Nature-2018} temperatures of the central floating reservoir $T_m\le 45~{\rm mK}$. Then bulk neutral excitations are thermally excited and lead to the thermal-metal-type behavior \cite{x38.5-tm}.

Points 1)-8) suggest that the mechanism \cite{x25-Simon} is not likely to  apply to the sample \cite{x19-Nature-2018}. The PH-Pfaffian hypothesis works better. Yet, it is important to look for other interpretations, and more research is necessary until the $5/2$ state is fully understood. New experimental  and numerical studies are needed, and it is crucial to reconcile experiment and numerics. Since numerics has strong record for the simplest filling factors, such as $1/3$, it is useful to address similarities and differences of $\nu=1/3$ and $\nu=5/2$. The quantum Hall effect at $\nu=1/3$ can be seen as the integer quantum Hall effect of composite fermions \cite{x39-book}. Weak disorder does not affect this physics qualitatively. Strong disorder is known to destroy the integer quantum Hall effect. There is experimental evidence \cite{x40-composite,xnew-composite} for composite fermions at $\nu=5/2$ too. The quantized plateau likely emerges due to their Cooper pairing \cite{x41-RG}. Multiple pairing channels exist \cite{x41.5-numerics} and disorder may affect them in a nontrivial way \cite{x42-super}. One possible mechanism was addressed in Refs. \onlinecite{x44-d2,x43-d1,x45-d3}. We will discuss another mechanism elsewhere.

\begin{acknowledgments}

The author thanks M. Heiblum and S. H. Simon for useful discussions and K. W. Ma for the critical reading of the manuscript. This research was supported in part by the National Science Foundation under Grant No. DMR-1607451.
\end{acknowledgments}


\begin{thebibliography}{99}
\bibitem{footnote2} Note that the correlation length of $W(x)$ may exceed the correlation length of the impurity potential.
\bibitem{xnew-foot2}
We do not expect possible edge reconstruction to significantly affect our estimates.
\bibitem{xnew-foot3}
Even weak nonrandom tunneling is unlikely to be negligible at small $\Delta k$ since it is likely to be relevant in the renormalization group sense.
\end{thebibliography}
\end{document}